\documentclass[useAMS,usenatbib,usegraphicx]{mn2e}

\usepackage{amsmath}
\usepackage{amssymb}
\usepackage[dvips]{color} 
\usepackage{gensymb}

\newcommand{\Msun}{\ensuremath{\mathrm{M_\odot}}}
\newcommand{\Rsun}{\ensuremath{\mathrm{R_\odot}}}
\newcommand{\Ni}{\ensuremath{^{56}\mathrm{Ni}}}
\newcommand{\Co}{\ensuremath{^{56}\mathrm{Co}}}
\newcommand{\Fe}{\ensuremath{^{56}\mathrm{Fe}}}
\newcommand{\Menv}{\ensuremath{M_\mathrm{env}}}
\newcommand{\Eexp}{\ensuremath{E_\mathrm{exp}}}
\newcommand{\Mni}{\ensuremath{M_\mathrm{\Ni}}}
\newcommand{\Mzams}{\ensuremath{M_\mathrm{ZAMS}}}

\title[Rapidly fading Type II supernovae]
{
On the nature of rapidly fading Type II supernovae
}
\author[T. J. Moriya et al.]
{Takashi J. Moriya$^1$\thanks{moriyatk@astro.uni-bonn.de},
 Maria V. Pruzhinskaya$^2$,
 Mattias Ergon$^3$, and \newauthor
 Sergei I. Blinnikov$^{4,5,6}$ \\
 $^1$
Argelander Institute for Astronomy, University of Bonn, Auf dem H\"ugel
71, D-53121 Bonn, Germany \\
$^2$
Lomonosov Moscow State University, Sternberg Astronomical Institute,
Universitetsky~pr.,~13,~Moscow, 119991, Russia \\
$^3$
The Oskar Klein Centre, Department of Astronomy, AlbaNova,
Stockholm University, SE-10691 Stockholm, Sweden \\
$^4$
Institute for Theoretical and Experimental Physics, Bolshaya
Cheremushkinskaya ulitsa 25, 117218 Moscow, Russia \\
$^5$
All-Russia Research Institute of Automatics,
Sushchevskaya ulitsa 22, 127055 Moscow, Russia \\
$^6$
Kavli Institute for the Physics and Mathematics of the
Universe (WPI), The University of Tokyo Institutes for Advanced Study, \\
The University of Tokyo, 
5-1-5 Kashiwanoha, Kashiwa, 277-8583 Chiba, Japan
}

\begin{document}

\date{Accepted 2015 October 06.  Received 2015 October 01; in original form 2015 May 26.}

\pagerange{\pageref{firstpage}--\pageref{lastpage}} \pubyear{2015}

\maketitle

\label{firstpage}

\begin{abstract}
It has been suggested that Type~II supernovae with rapidly fading
light curves (a.k.a. Type~IIL supernovae) are explosions
of progenitors with low-mass hydrogen-rich envelopes
which are of the order of 1~\Msun.
We investigate light-curve properties of supernovae from such progenitors.
We confirm that such progenitors
lead to rapidly fading Type~II supernovae.
We find that the luminosity of supernovae from such progenitors
with the canonical explosion energy of $10^{51}$~erg and \Ni\ mass of
0.05~\Msun\ can increase temporarily
shortly before all the hydrogen in the envelope recombines.
As a result, a bump appears in their light curves.
The bump appears because the heating from the nuclear decay of \Ni\
can keep the bottom of hydrogen-rich layers in the ejecta ionized,
and thus the photosphere can stay there for a while.
We find that the light-curve bump becomes less significant when
we make explosion energy larger ($\gtrsim 2\times 10^{51}$~erg),
\Ni\ mass smaller ($\lesssim 0.01$~\Msun), \Ni\ mixed in the ejecta,
or the progenitor radius larger.
Helium mixing in hydrogen-rich layers makes the light-curve decline
 rates large but does not help reducing the light-curve bump.
Because the light-curve bump we found in our light-curve models
has not been observed in rapidly fading Type~II supernovae,
they may be characterized by not only low-mass hydrogen-rich envelopes
but also higher explosion energy,
larger degrees of \Ni\ mixing,
and/or larger progenitor radii
than slowly fading Type~II supernovae,
so that the light-curve bump does not become significant.
\end{abstract}


\section{Introduction}
Type~II supernovae (SNe~II) are SNe which show hydrogen features in
their spectra (e.g., \citealt{filippenko1997}).
It has been believed that they are
explosions of massive stars which have kept
their hydrogen-rich envelope until the time of the core collapse.
Many direct detections of SN II progenitors, including
the blue supergiant star in Sanduleak~-69\degree 202 exploded as SN~1987A,
prove this idea (see \citealt{smartt2009} for a review).

\begin{table*}
\caption{Progenitor properties.}
\label{table:pp}
\begin{center}         
\begin{tabular}{cccccccc}
\hline       
  $M_\mathrm{ZAMS}$ & $f$ & $M_\mathrm{fin}^a$ & 
  $M_\mathrm{env}$ & $X_\mathrm{surf}^b$ &
  $R$ & $L^c$ & $T_\mathrm{eff}^d$\\ 
  $\Msun$ &  & $\Msun$ & 
  $\Msun$ &  &
  $10^3 \Rsun$ & $10^6 \mathrm{L_\odot}$ & $10^3$K\\ 
\hline                    
15 & 1.0 & 13.9 & 10.2 & 0.66 & 0.65 & 0.52 & 3.42 \\
20 & 2.5 & 10.5 & 4.4  & 0.65 & 1.10 & 1.23 & 3.27 \\
20 & 3.0 & 8.0  & 2.0  & 0.62 & 0.89 & 1.10 & 3.09 \\
25 & 1.0 & 14.0 & 4.1  & 0.67 & 1.10 & 2.26 & 3.81 \\
25 & 1.5 & 10.6 & 1.2  & 0.41 & 0.70 & 1.93 & 4.57 \\
\hline
\end{tabular}
\end{center}
{\footnotesize
$^a$ Total mass, 
$^b$ Surface hydrogen abundance,
$^c$ Luminosity, 
$^d$ Effective temperature
}
\end{table*}

Even though SNe~II commonly show hydrogen features in their spectra,
their light-curve (LC) evolution is heterogeneous\footnote{
We do not consider SNe~IIn in this paper because their LCs are
strongly affected by dense circumstellar media (e.g., \citealt{moriya2013}).}.
Classically, SN~II LCs are classified in two distinct sub-classes,
SNe~IIP and IIL, based on their LCs.
SNe~IIP show a LC ``plateau'' and their luminosity does not change much
for about 100 days. On the contrary, the luminosity of SNe~IIL
constantly declines and their LCs look ``linear''.
While some studies suggest that these two sub-classes are
distinct from each other \citep{arcavi2012,faran2014a,faran2014b,poznanski2015},
others indicate that SNe~IIP and IIL are not clearly separated
and SN~II LC decline rates have a continuous distribution \citep{anderson2014,sanders2014}.
In this paper, we continue to use the word ``SN~IIL'' to naively refer to
SNe~II whose LCs decline relatively rapidly.
We note that it is not our intention to suggest that SNe~IIL are a distinct
class of SNe~II when we say SNe~IIL, nor to define SNe~IIL.
Some SNe~II which we call SNe~IIL in this paper may
be classified as SNe~IIP depending on how they are separated.

The primary origin of the diversity in the LC decline rates of SNe~II 
are presumed to be in the remaining amount of hydrogen in the envelope
(e.g., \citealt{grassberg1971,falk1977,litvinova1983,swartz1991},
see also \citealt{blinnikov1993a,young2005}).
The luminosity of SNe~II with less hydrogen-rich envelope masses
declines more rapidly because of the low ejecta density (see also \citealt{goldfriend2014,pejcha2014}).
If the hydrogen-rich envelope mass (\Menv) is sufficiently small,
the SNe are observed as SNe~IIb
in which the spectral features transit from those of SNe~II to SNe~Ib
(see \citealt{bufano2014} for a recent collection of SNe~IIb).
LC modeling of observed SNe indicates that
the hydrogen-rich envelope masses of SN~IIP progenitors
are $\sim 5-15~\Msun$
(e.g., \citealt{popov1993,bersten2011,inserra2013,utrobin2013,pejcha2015}),
while those of SN~IIb progenitors are
less than $\sim 0.5~\Msun$
(e.g., \citealt{shigeyama1994,woosley1994,blinnikov1998,bersten2012,ergon2014}).

Most of SN~II LC studies based on theoretical SN progenitors
from stellar evolution 
have been performed for the progenitors
with $\Menv\gtrsim 5$~\Msun\ which become SNe~IIP
(e.g., \citealt{kasen2009,dessart2010,dessart2011b,dessart2013,utrobin2014}),
and those with $\Menv\lesssim 0.5~\Msun$ which become SNe~IIb
(e.g., \citealt{shigeyama1994,woosley1994,dessart2011,dessart2015}).
There exist only a few LC studies for the SN~II progenitors 
with $\Menv\sim 1~\Msun$ which are presumed to become rapidly fading SNe~II
or SNe~IIL \citep{blinnikov1993a,baklanov2002,young2004,dessart2010,morozova2015}.
In this paper, we focus on SN~II progenitors with a hydrogen-rich
envelope mass of $\sim 1~\Msun$ and perform a numerical LC study
of such SN~II progenitors.
We investigate the nature of rapidly fading SNe~II
by numerically modeling LCs from such progenitors.

The rest of this paper is organized as follows.
We introduce our progenitor models with $\Menv\sim 1~\Msun$ in
Section~\ref{sec:progenitor}.
We show LC models from our progenitors in Section~\ref{sec:lightcurve}.
We discuss the nature of rapidly fading SNe~II based on the obtained LC models
in Section~\ref{sec:discussion}.
We conclude this paper in Section~\ref{sec:conclusion}.

\section{Progenitors with low-mass hydrogen-rich envelopes}\label{sec:progenitor}
We use a public stellar evolution code \texttt{MESAstar}
version 6794 \citep{paxton2011,paxton2013,paxton2015}
to obtain SN progenitors with low-mass hydrogen-rich envelopes.
We evolve stars with the zero-age main-sequence (ZAMS) masses ($M_\mathrm{ZAMS}$)
of 15, 20, and 25 \Msun. The initial metallicity is assumed to be
$Z=0.02$.
The stellar evolution is followed until the core collapse, except
for one model specified below.
Convection is treated with the mixing-length theory
with a mixing-length parameter of $1.6$.
The Ledoux criterion for convection is adopted
with a semiconvective efficiency of $1.0$.
No overshooting is taken into account.

We obtain several SN progenitors with low-mass hydrogen-rich envelopes
by changing their mass-loss rates.
The standard mass-loss rates $\dot{M}_\mathrm{st}$ are
the ``Dutch'' mass-loss rates adopted in \texttt{MESAstar}.
The standard mass-loss rates for stars with the effective temperature
higher than $10^4$~K are based on \citet{vink2001}. 
Those for stars cooler than $10^4$~K are from \citet{dejager1988}.
The actual mass-loss rates ($\dot{M}$) used in our calculations are
\begin{equation}
 \dot{M}=f\dot{M}_\mathrm{st}.
\end{equation}
The factor $f$ is made as large as 3 in this study.
The observed mass-loss rates scatter more than by a factor of 3,
and the adopted mass-loss rates are within uncertainties
(e.g., \citealt{meynet2014}).

Table \ref{table:pp} summarizes the properties of the SN progenitors
we have numerically evolved for this study.
We obtain four stars with \Menv\ between 1 and 5~\Msun.
We also obtain a star with $\Menv=10.2~\Msun$
($M_\mathrm{ZAMS}=15~\Msun$) for comparison.
The evolution of the star with $M_\mathrm{ZAMS}=20~\Msun$ and $f=3.0$
is only followed up to the end of the core Ne burning.
However, the star is expected to explode within several years
after the core Ne burning and the envelope structure and the final
mass are not likely to be affected in the remaining time to
the core collapse.
The other stars are evolved until the core collapse.

\begin{figure*}
 \begin{center}
  \includegraphics[width=\columnwidth]{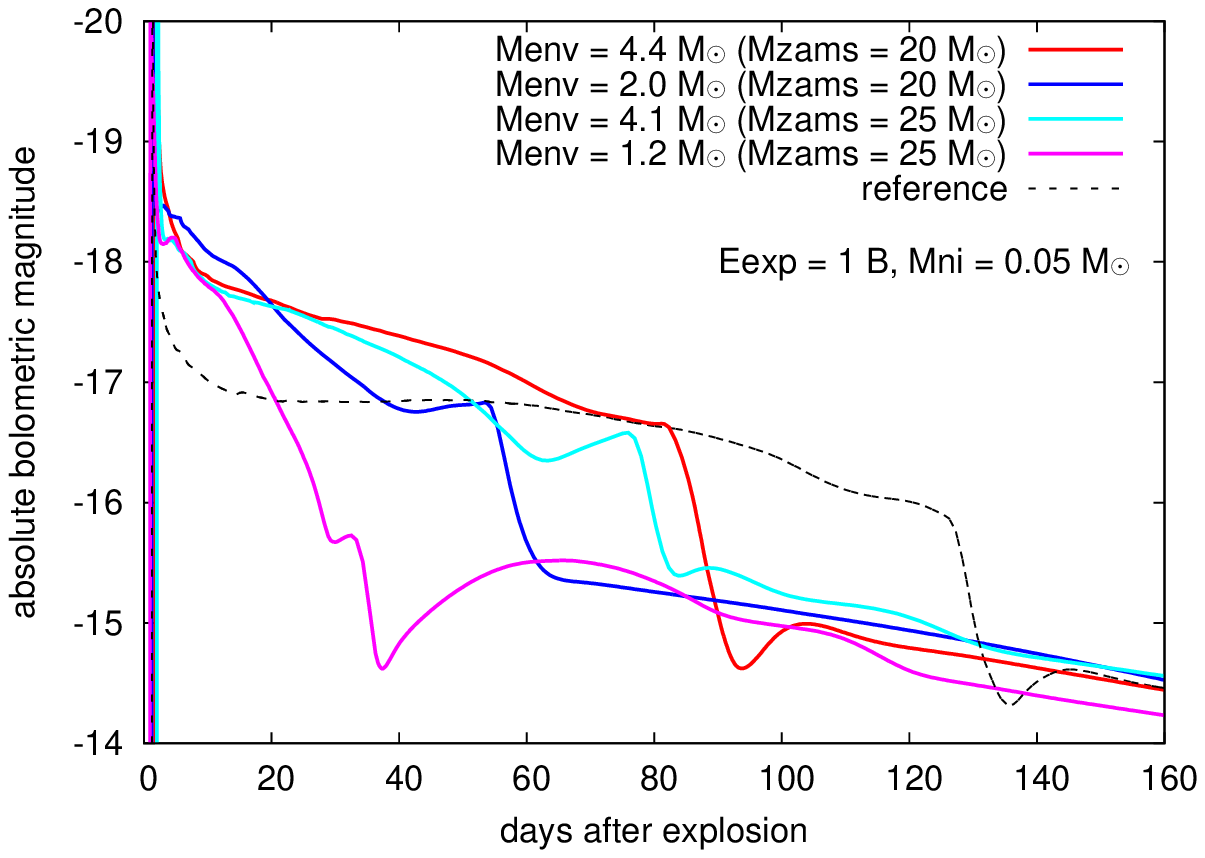} 
  \includegraphics[width=\columnwidth]{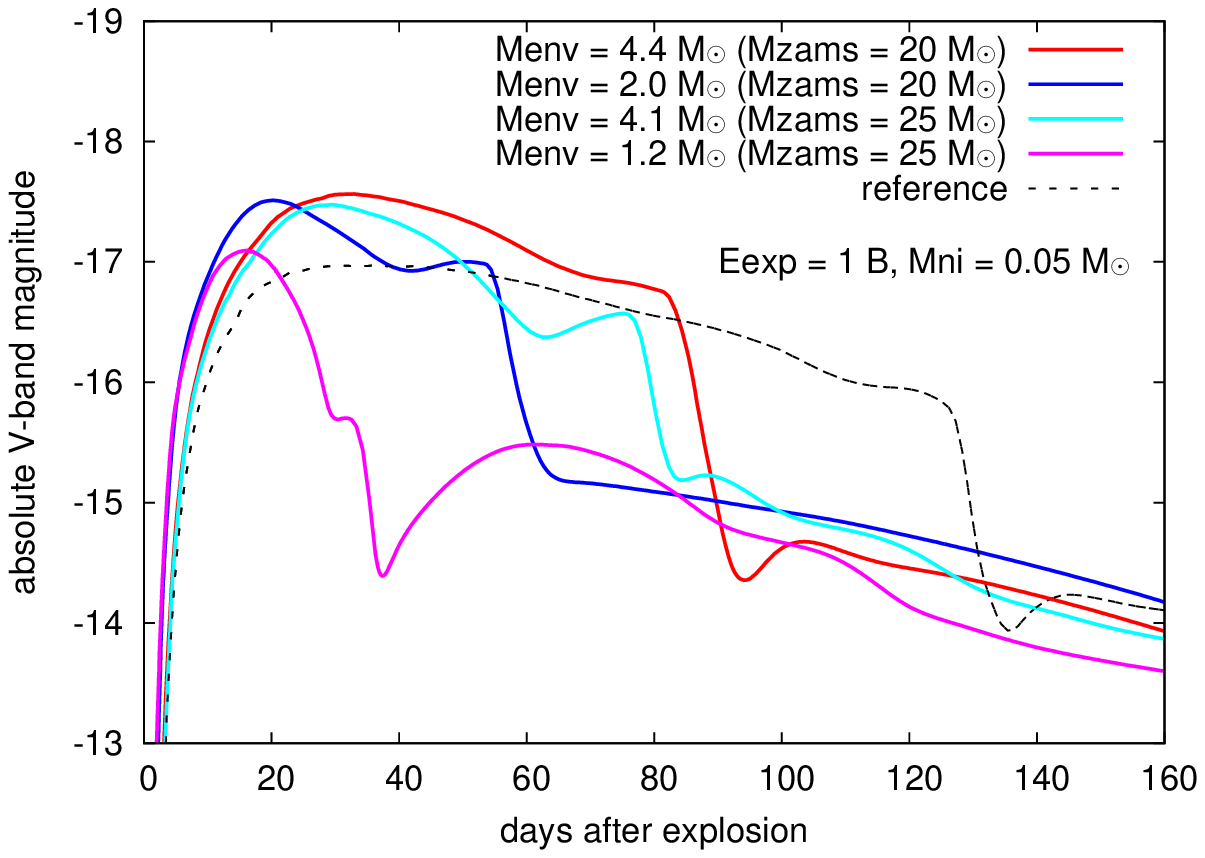} 
 \end{center}
\caption{
Bolometric (left) and $V$-band (right) LCs from
our progenitors with low-mass hydrogen-rich envelopes
exploded with the canonical explosion energy (1~B) and \Ni\ mass (0.05~\Msun).
The reference LC are from our 15~\Msun\ progenitor model.
}\label{fig:standard}
\end{figure*}

\section{Light curves}\label{sec:lightcurve}
We perform numerical LC calculations for
the SN progenitors described in the previous section.
We use a one-dimensional multi-frequency radiation hydrodynamics code
\texttt{STELLA} to follow LCs numerically (e.g., \citealt{blinnikov1993a,blinnikov1998,blinnikov2006}).
We take the central 1.4 \Msun\ of the SN progenitors away for the LC
calculations, assuming that the central region becomes a neutron star.
SN explosions are initiated by putting thermal energy to the innermost
layers.
We do not follow the explosive nucleosynthesis. Thus,
the SN ejecta composition is the same as the pre-SN composition
except for \Ni.
When we do not take mixing into account, \Ni\ is located at the center
of SN ejecta. When we mix \Ni\ (Section~\ref{sec:mixing}),
\Ni\ is evenly mixed within the mixed region.
We also investigate the effect of helium mixing into the hydrogen-rich
envelope (Section~\ref{sec:hemixing}).
As our focus in this paper is on the early LCs when the photosphere is
located in the hydrogen-rich envelopes,
the effect of the explosive nucleosynthesis on our results
is presumed to be small.

\begin{figure}
 \begin{center}
  \includegraphics[width=\columnwidth]{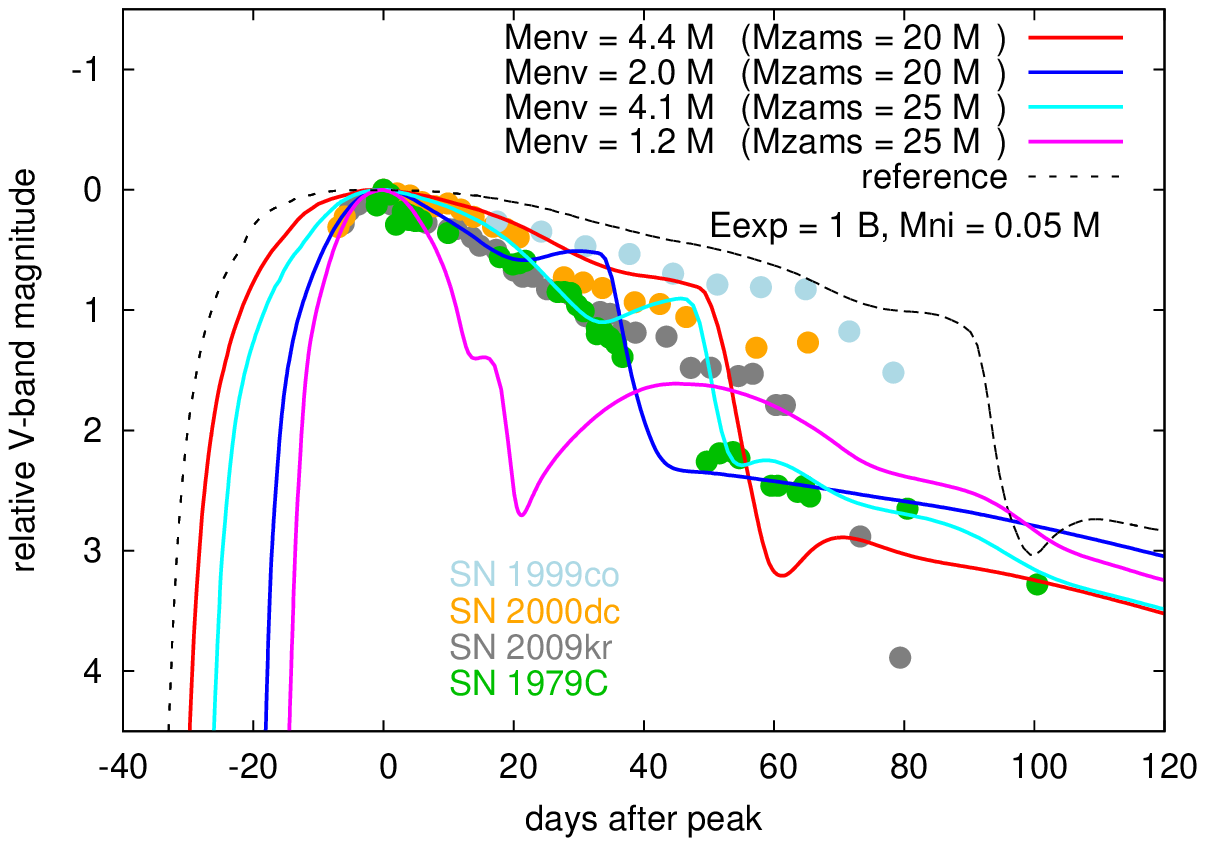} 
 \end{center}
\caption{
$V$-band LCs of our canonical explosion models scaled at the LC peak.
The $V$-band LCs of SNe~II with several LC decline rates are
also presented.
The observational data are from \citet{faran2014b} (SNe~1999co and 2000dc),
\citet{elias-rosa2010} (SN~2009kr), and \citet{devaucouleurs1981} (SN~1979C).
SNe~2009kr and 1979C are among the most rapidly fading SNe~II
\citep[e.g.,][]{faran2014b,anderson2014}.
}\label{fig:standardVscaled}
\end{figure}

\subsection{Canonical explosion energy and \Ni\ mass}\label{sec:canonical}
We adopt an explosion energy of $10^{51}\ \mathrm{erg}\equiv 1\ \mathrm{B}$
and a \Ni\ mass of 0.05 \Msun\ as the canonical values based on
the observations of SNe~II (e.g., \citealt{hamuy2003,anderson2014,sanders2014}).
\Ni\ is located at the center of the ejecta in the canonical models,
although SNe~II are known to be affected by mixing \citep[e.g.,][]{arnett1989}.
The effect of mixing is discussed in the later sections.

Figure~\ref{fig:standard} presents our LCs with the canonical explosion
parameters.
The reference LC is from our $M_\mathrm{ZAMS}=15~\Msun$ model which
has $\Menv = 10.2~\Msun$. The bolometric luminosity does not change
significantly for more than 50~days when the recombination wave travels
in the hydrogen-rich envelope.
Then, the LC drops when the recombination wave disappears from the
hydrogen-rich envelope.
The $V$-band LC
decline rate after the peak before the drop is 0.48 $\mathrm{mag~day^{-1}}$.
We see rebrightening after the LC drop before
the LC start to follow the nuclear decay.
In this phase, the SN have already entered the nebular phase.
Our code is not suitable to follow the nebular phase and this
rebrightening may be a numerical artifact.

The early luminosity in the LC models with $\Menv\sim 1~\Msun$
is brighter than that of the reference LC
because of the lower envelope masses \citep[e.g.,][]{kasen2009}.
In Fig.~\ref{fig:standardVscaled},
we show the $V$-band LCs scaled with their peak magnitudes.
We can find that SN~II progenitors with low-mass
hydrogen-rich envelopes result in rapidly fading SNe~II as
expected from previous studies.
We also show the observed $V$-band LCs for SNe~II with relatively
large decline rates.
We can see that our low-mass envelope progenitors provide
LC decline rates consistent with the observations.
Especially, the $V$-band LC decline rates for our models with
$M_\mathrm{env}=2.0$~\Msun\ and 4.1~\Msun\ are consistent
with those of SNe~2009kr and 1979C, which
are among the most rapidly fading SNe~II and often referred as SNe~IIL.

Looking into the LC of SN~2009kr as an example,
a substantial LC drop is observed at around 60 days after the peak.
This kind of the LC drop is observed in several rapidly fading SNe
(e.g., \citealt{valenti2015}).
The LC drop caused by the recombination wave reaching the bottom of the
hydrogen-rich envelope in the model with $\Menv=4.1~\Msun$ is
found at around 50~days after the peak and there is only
a difference of about 10~days.

A significant difference between our numerical LCs and observed ones is
in the existence of a period of a slight luminosity increase between
the early fading phase and the LC drop.
The bolometric luminosity increase is found in three of our canonical models.
The luminosity increase is also found in the $V$-band LCs, 
although the increase in the $V$-band LC with $\Menv=1.2~\Msun$
is less significant and it is almost flat.
The remaining LC with $\Menv=4.4~\Msun$ does not show the luminosity
increase, but the LC flattens before the drop.
We call the increasing part of the LCs we found in our LC models
as a ``bump'' in the following.
We clearly see the LC bump in the models with $\Menv=4.1$ and 2.0~\Msun,
while it is vague in the model with $\Menv=4.4~\Msun$.
The model with $\Menv=1.2~\Msun$ shows the bump in a short timescale.
Thus, we are likely to observe the LC bump for a long time
in the canonical explosions of the progenitors with $\Menv\simeq
2-4~\Msun$.
This kind of bumps are also found in other numerical LC models of
rapidly fading SNe~II (e.g., \citealt{swartz1991,baklanov2002,young2004}),
but they are not paid attention much.
The origin of the bump is discussed in the next section
where LC models with different \Ni\ masses and explosion energies
are presented.
In this study, we focus on this bump 
and try to use it to deduce the nature of rapidly fading SNe~II.

Another LC feature to note is in our LC model of the smallest
\Menv\ ($\Menv = 1.2~\Msun$).
Because the recombination wave recedes in the hydrogen-rich envelope 
very quickly due to the small \Menv,
the early LC declines in a very short timescale.
Then, after the LC drop following the bump discussed above,
the luminosity increases again instead of following the \Co\ nuclear
decay rate. The peak time of this part of the LC ($\sim$ 60 days)
corresponds to the diffusion time in the core, and thus
the luminosity increase is likely due to the diffusion in the core.
The hydrogen-rich envelope mass in the model is close to those in
SNe~IIb and the LC starts to be similar to those of SNe~IIb
with $\Menv=1.2~\Msun$.
The bolometric LC shape is also somewhat similar to that of SN~1987A,
although our LC evolves in shorter timescales (e.g., \citealt{suntzeff1992}).
This kind of SNe~II may exist
as an intermediate type between SNe~IIL and SNe~IIb.

\begin{figure}
 \begin{center}
  \includegraphics[width=\columnwidth]{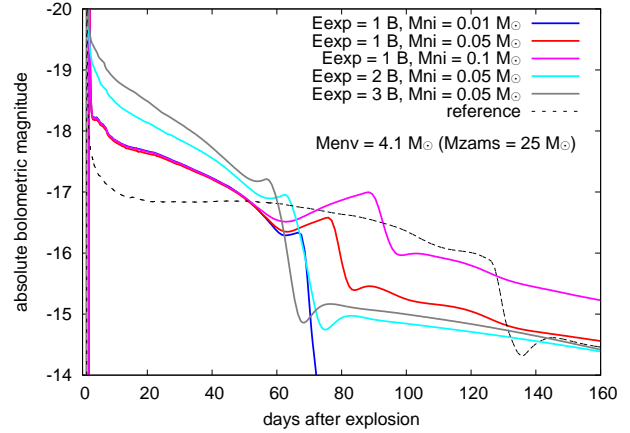} 
 \end{center}
\caption{
Bolometric LCs from the explosions of our progenitor model
with $\Menv=4.1~\Msun$ with different explosion energies and \Ni\ masses.
The reference LC is the same as in Fig.~\ref{fig:standard}.
}\label{fig:diffeni}
\end{figure}

\begin{figure}
 \begin{center}
  \includegraphics[width=\columnwidth]{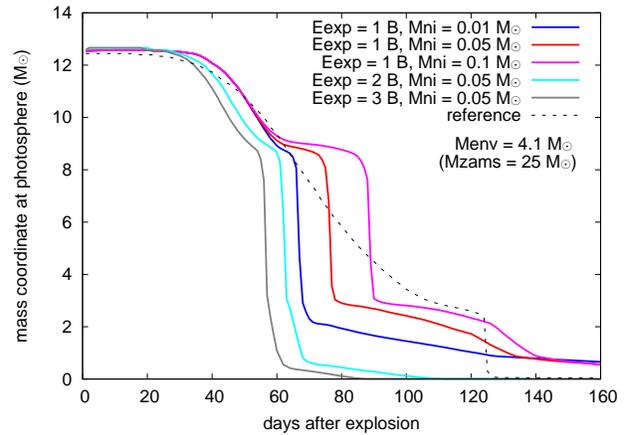} 
 \end{center}
\caption{
Evolution of the location of photosphere in our LC models presented in
Fig.~\ref{fig:diffeni}.
}\label{fig:photomass}
\end{figure}

\subsection{Other explosion energies and \Ni\ masses}
To investigate the cause of the LC bump found in our rapidly fading
SN~II models, we explore LCs with different explosion energies and
\Ni\ masses in this section. 
For this purpose, we use the 14.0 \Msun\ progenitor model
with $\Menv=4.1\ \Msun$ ($\Mzams=25~\Msun$).
We choose this model for our further investigation because
(i) the LC from the canonical explosion parameters has
a significant bump,
(ii) the progenitor has almost the same mass as our
reference model (14 \Msun),
and
(iii)
the progenitor is obtained without modifying the mass-loss rates $(f=1.0)$.
General bump properties found for this progenitor model
with different explosion energies and \Ni\ masses are
also found in other progenitors.
We use
three different explosion energies (1, 2, and 3 B)
and
\Ni\ masses (0.01, 0.05, and 0.1 \Msun).

Figure~\ref{fig:diffeni} shows the LCs with different 
explosion energies and \Ni\ masses.
First of all, we find that the bump is strongly affected by the \Ni\ mass.
When the \Ni\ mass is increased to 0.1 \Msun, the duration of the bump
becomes twice as long as that of the model with $\Mni=0.05~\Msun$.
When the \Ni\ mass is decreased to 0.01 \Msun, the LC bump almost
disappears and there only remains a brief almost plateau phase after the
early declining phase.
In the case of SNe~IIP, the amount of \Ni\ affects the duration of the
plateau phase, but we do not expect the formation of the bump
in this \Ni\ mass range \citep{kasen2009}.
\citet{dessart2010} presented a LC model with
$\Menv=2.2~\Msun$, but they did not include \Ni.
This may be why they did not find the bump in their LC.

The fact that the LC bump is strongly affected by the \Ni\ mass
indicates that the LC bump is caused by the heating from the nuclear
decay of $\Ni\rightarrow\Co\rightarrow\Fe$.
In Fig.~\ref{fig:photomass},
we show the location of the photosphere in the models presented in
Fig.~\ref{fig:diffeni}.
Photosphere is defined as the location where the Rosseland mean
optical depth becomes $2/3$.
The photosphere recedes inward similarly at first
in the models with different \Ni\ masses.
Then, when the photosphere reaches near the bottom of the hydrogen-rich
envelope, the photosphere stays there for a while because
the heating due to the nuclear decay can keep hydrogen ionized.
Because the radius of the photosphere increases with time
if it stays in the same mass coordinate, the luminosity increases
during this epoch. Thus, the bump appears in the LCs.
The photosphere eventually goes into the core when hydrogen
in the envelope can no longer be kept ionized, and the LCs drop at that moment.
Because larger amount of \Ni\ can keep hydrogen ionized longer,
the bump appears more significantly with larger amount of \Ni.

Next, we investigate the effect of explosion energy.
Figure~\ref{fig:diffeni} shows LCs with different explosion energies.
The evolution of the location of the photosphere is in Fig.~\ref{fig:photomass}.
The early luminosity becomes larger as the explosion energy increases.
Meanwhile, the bump is significantly reduced or almost disappears when
the explosion energy is above 2 B when $\Mni=0.05~\Msun$.
This is because more rapid expansion results in more rapid reduction
of the gamma-ray optical depth, and the heating from the nuclear decay
is less effective.
In addition, the adiabatic cooling in the SN ejecta 
becomes more effective with higher explosion energy.
As a result,
the photosphere no longer stays at the bottom of the hydrogen-rich layers for a
long time.
The LC models in \citet{blinnikov1993a} do not have significant
bumps presumably because they mostly present LCs for an explosion
energy of 2~B.

To summarize, we find that large explosion energies ($\Eexp\gtrsim 2$~B)
and small \Ni\ masses ($\Mni\lesssim 0.01~\Msun$) make the LC bumps
less significant.

\begin{figure}
 \begin{center}
  \includegraphics[width=\columnwidth]{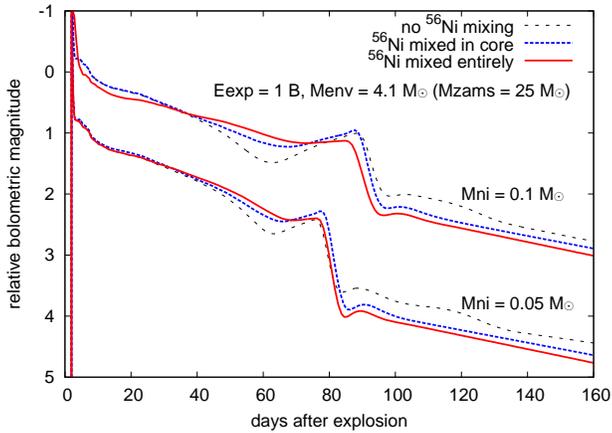} 
 \end{center}
\caption{
Bolometric LCs with different degrees of \Ni\ mixing.
The effect of the mixing is explored with the explosions of
the $\Menv=4.1~\Msun$ progenitor with the canonical explosion energy (1~B)
and two different \Ni\ masses (0.05 \Msun\ and 0.1 \Msun).
The models with the same \Ni\ mass are scaled with the same magnitude.
}\label{fig:mixing}
\end{figure}

\begin{figure}
 \begin{center}
  \includegraphics[width=\columnwidth]{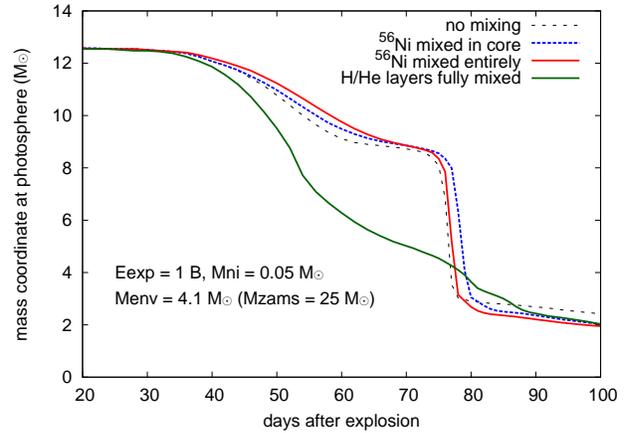} 
 \end{center}
\caption{
Evolution of the location of photosphere in the LC models with different
mixing presented in Figs.~\ref{fig:mixing} and \ref{fig:hemixing}.
}\label{fig:photomix}
\end{figure}

\begin{figure}
 \begin{center}
  \includegraphics[width=\columnwidth]{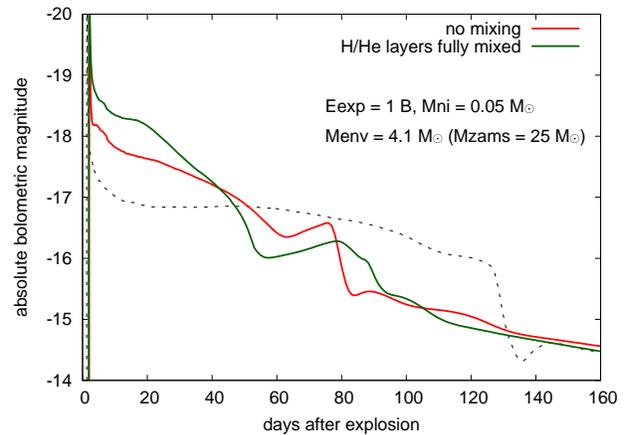} 
 \end{center}
\caption{
Bolometric LC from the progenitor in which hydrogen-rich and helium
 layers are fully mixed.
}\label{fig:hemixing}
\end{figure}

\subsection{\Ni\ mixing}\label{sec:mixing}
In all the models presented so far, \Ni\ is placed at the center of SN ejecta.
We find that the heating from the nuclear decay of \Ni\ strongly affects
the LC bump in the previous section.
Since the heating from the nuclear decay depends on
the degree of the \Ni\ mixing, we investigate the effect of the \Ni\ mixing
on the LC bumps.
We assume two kinds of \Ni\ mixing in our LC calculations.
\Ni\ is mixed only in the core in one case and
in the entire SN ejecta in the other.
\Ni\ is uniformly mixed in both cases.

Figure~\ref{fig:mixing} shows how the LC bump is affected by the \Ni\ mixing.
The higher degree of mixing makes the LC bump less significant.
This is because the photosphere recedes more slowly
with the \Ni\ mixing thanks to the more efficient heating by the nuclear decay
in the outer hydrogen-rich layers.
The slower recession of the photosphere
can be seen in Fig.~\ref{fig:photomix} where the evolution of
photosphere in mass coordinate is shown.

Another interesting consequence of the \Ni\ mixing appears in the late-time
LC tails after the luminosity drop.
All the late-time LCs decline with the timescale of the \Co\ decay.
Meanwhile, the luminosity of the tails depends on the degree of the \Ni\ mixing.
This is simply because gamma-rays from the nuclear decay are less likely
to be absorbed by the ejecta if they are emitted in outer layers.
Recently, \citet{wheeler2014} argued that the LC tails
of Type IIb/Ib/Ic SNe have much diversity even if their early LCs
have similar properties. They ascribed the diversity on
the explosion asymmetry. However, the difference in the degree of
the \Ni\ mixing, which is not considered in the analytic LC model of
\citet{arnett1982} they use, can also make the diversity in the
late-time LCs \citep[see also][]{ergon2014}.

\begin{figure}
 \begin{center}
  \includegraphics[width=\columnwidth]{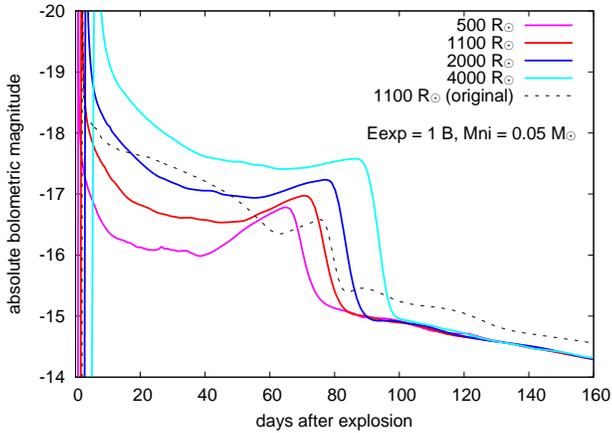} 
 \end{center}
\caption{
Bolometric LCs from simplified progenitors with different radii.
We also show the original model with the same mass 
($M_\mathrm{fin}=14.0~\Msun$ and $\Menv=4.1~\Msun$)
whose radius is 1100~\Rsun.
}\label{fig:radii}
\end{figure}

\subsection{Helium mixing}\label{sec:hemixing}
Multi-dimensional simulations of stellar explosions show that
the hydrogen-rich layer and helium core are mixed during the explosion
because of the Rayleigh-Taylor instability \citep[e.g.,][]{hachisu1990}.
The mixing of helium into the hydrogen-rich layers affects the LC shape
\citep[e.g.,][]{woosley1988,shigeyama1990,utrobin2007}.
It is also suggested that the helium mixing in the hydrogen-rich layers
can result in the rapid declines in SNe~II \citep[e.g.,][]{swartz1991}.
We investigate the effect of the helium mixing in the hydrogen-rich
layer on the LC bump.

Our progenitor model with $\Menv=4.1~\Msun$ has helium mass of
6.4~\Msun\ and hydrogen mass of 3.1~\Msun\ in total
in hydrogen-rich and helium layers. Based on this progenitor model,
we construct a progenitor model
in which the hydrogen-rich and helium layers are fully mixed.
Thus, the model has 9.5~\Msun\ of the mixed outer layer with the hydrogen
fraction of 0.33 and the helium fraction of 0.67.

The LC from the mixed progenitor is presented in Fig.~\ref{fig:hemixing}.
As is found in previous studies, the higher helium fraction in the
envelope due to the helium mixing makes the early LC brighter
\citep[e.g.,][]{swartz1991,kasen2009}.
The higher helium fraction makes the recombination temperature higher, and
it results in the fast recession of the recombination front
(Fig.~\ref{fig:photomix}).
However, the heating from the nuclear decay can still keep the innermost
layers of the mixed envelope ionized.
Thus, the helium mixing does not result in the
reduction nor disappearance of the LC bump.
The duration of the LC bump with the helium mixing
actually gets longer than that without the helium
mixing, because the photosphere travels faster than in the model without
mixing and it reaches the bottom of the mixed layer earlier.

\subsection{Progenitor radii}\label{sec:radii}
Radii of SN~II progenitors change their LC shapes (e.g., \citealt{young2004}).
The LC models presented so far are based on the progenitors developed
with \texttt{MESAstar}, and thus our progenitor radii are based on
the stellar structure obtained by the code.
However, there are uncertainties in the stellar evolution theory, and
the stellar radii obtained by the code have uncertainties.
\citet{dessart2013} suggested that SNe~IIP progenitors may have smaller
radii than those predicted by the stellar evolution theory based on
their LC models (see also \citealt{gonzalez2015}),
while violent nuclear burning at the stellar core in the final stages
of the stellar evolution may result in the stellar surface expansion
(e.g., \citealt{quataert2012,mcley2014,moriya2014}).
We investigate how the uncertainties in the progenitor radii
affect the LC bump in this section.

We use simplified stellar structure to obtain SN~II progenitors
with different radii.
We keep the progenitor and envelope mass the same as
the original model with $M_\mathrm{fin}=14.0~\Msun$ and
$\Menv=4.1~\Msun$ developed from
$\Mzams=25~\Msun$ which has the progenitor radius $R=1100~\Rsun$.
The stellar structure is constructed based on the polytropic equation
of state with the polytropic index of 3 (see, e.g.,
\citealt{blinnikov1993a,baklanov2002} for details).
The stellar interior abundance is kept the same as in the original model.
We explode the progenitor with the canonical explosion energy (1~B)
and \Ni\ mass (0.05~\Msun) without any mixing.

In Fig.~\ref{fig:radii}, we show the LC models from the progenitors
with different radii and the original model from \texttt{MESAstar}. 
Although the original model and the artificial model with the same radius
($1100~\Rsun$) have different early LC evolution because of the
difference in the density structure, the LC bump is still found in
both models. The LC bump becomes less significant in larger progenitors.
This is because the early luminosity during the plateau phase is
brighter when the progenitor is larger
(e.g., \citealt{kasen2009}).
The effect of the temporal
luminosity increase is less significant because of
the higher early luminosity caused by the larger progenitor radii.

\section{Discussion}\label{sec:discussion}
\subsection{Photospheric velocity}
We have focused on the LC properties from the explosions of the progenitors with
hydrogen-rich envelope mass of $\sim 1~\Msun$. For the completeness,
we show the photospheric velocities of the models we have presented so
far. Photospheric velocities affect the spectral properties of SNe, and
the information on photospheric velocities is complementary to the LC properties.

\begin{figure}
 \begin{center}
  \includegraphics[width=\columnwidth]{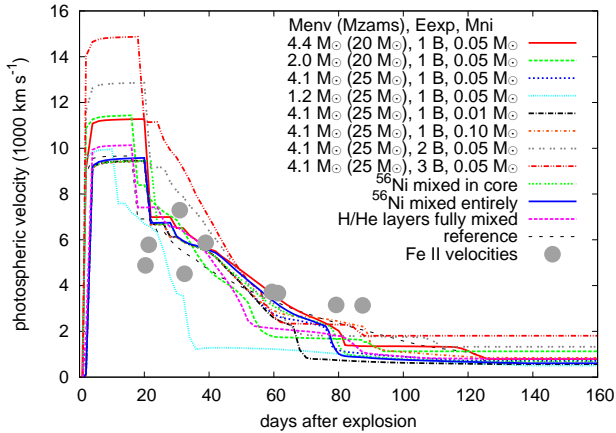} 
 \end{center}
\caption{
Photospheric velocities of the explosion models presented in this study.
The collection of Fe II velocities observed in rapidly fading SNe~II
presented in \citet{faran2014b} scaled with a typical observed
 photospheric velocity at 50 days
(4000~$\mathrm{km~s^{-1}}$, e.g., \citealt{poznanski2015})
is shown for comparison.
}\label{fig:photov}
\end{figure}

Figure~\ref{fig:photov} shows the photospheric velocities of our models.
They are within those estimated in rapidly fading SNe~II
\citep[e.g.,][]{faran2014b}. 
Our rapidly fading SN~II models typically have higher photospheric
velocities than the reference model, as is observed in rapidly fading
SNe~II \citep[e.g.,][]{faran2014b}.
This is because of the lower envelope mass in the rapidly fading SNe~II.
The photospheric velocities are mainly affected by the explosion energy
and the helium mixing in a given progenitor model.

\subsection{Nature of rapidly fading SNe~II indicated from the LC bump}
Based on the LC models obtained so far, we discuss the nature
of rapidly fading SNe~II.
We have confirmed that SNe~II from progenitors with low-mass
hydrogen-rich envelopes fade quickly, and their decline rates are
consistent with those observed in rapidly fading SNe~II.
However, our LC models with
the canonical explosion energy (\Eexp\ = 1~B) and \Ni\ mass
(\Mni\ = 0.05~\Msun) often show a temporal luminosity increase (LC bump)
before all the hydrogen in the hydrogen-rich envelope recombines.
This kind of LC bumps have not been observed in rapidly fading SNe~II.
This implies that rapidly fading SNe~II may not only be characterized by 
low-mass hydrogen-rich envelopes.
Some explosion properties which prevent LCs from having the LC bump
may be commonly shared among rapidly fading SNe~II.

We have found four possible ways to make the LC bump less significant,
i.e., large explosion energies ($\Eexp\gtrsim 2$~B),
small \Ni\ masses ($\Mni\lesssim 0.01$~\Msun), 
large degrees of \Ni\ mixing, and
large progenitor radii.
If SN~II progenitors with low-mass hydrogen-rich envelopes
tend to come from relatively more massive stars among SN~II progenitors,
they are likely to have larger explosion energy, according to
the empirical relation between explosion energy and progenitor mass
(e.g., \citealt{hamuy2003,nomoto2011,utrobin2013,poznanski2013,pejcha2015,bose2015}).
Then, the lack of the LC bump may indicate that most rapidly fading SNe~II
have low-mass hydrogen-rich envelopes because of higher ZAMS masses,
which end up with both low-mass hydrogen-rich envelopes and high
explosion energies.
SN~II progenitors with low-mass hydrogen-rich envelopes can also
originate from small ZAMS mass stars if they are in binary systems
thanks to the mass loss by binary interactions
(e.g., \citealt{nomoto1995,eldridge2008}).
However, small ZAMS mass SN~IIL progenitors are disfavored if SNe~IIL
typically have larger explosion energies than SNe~IIP,
contrary to SNe~IIb which may prefer small ZAMS mass progenitors in binary systems
(e.g., \citealt{benvenuto2013,folatelli2014,jerkstrand2015}).
There are other observational indications that SNe~IIL may prefer
higher ZAMS mass progenitors than SNe~IIP
(e.g., \citealt{elias-rosa2011,anderson2012,hanin2013},
see \citealt{maund2015} for the case of SN~2009kr).

A small amount of \Ni\ in the ejecta is also shown to reduce the LC bump.
However, SN~II observations indicate that rapidly fading SNe~II
tend to have slightly higher \Ni\ masses than slowly fading ones
\citep{anderson2014}. Thus, observations disfavor this way of making
the LC bump less significant, although several possible mechanisms
can make \Ni\ masses of SNe~II small.
For example, some high mass progenitors may suffer significantly from fallback,
resulting in small amount of \Ni\ ejection.

The LC bump is also shown to be reduced by large degrees of the \Ni\ mixing.
Some SNe~II are suggested to experience significant mixing in the SN
ejecta (e.g., SN~1987A, \citealt{arnett1989}). The lack of the LC bump in observed
SNe~II may thus indicate that SNe~II generally have large degrees of
internal mixing, dredging up \Ni\ above the stellar core.

It has been suggested that the rapidly fading SNe~II may result from
the larger degree of helium mixing \citep[e.g.,][]{swartz1991}.
We have confirmed that the helium mixing does make the LC decline rate
of SNe~II larger. However, the helium mixing does not result in the
reduction of the LC bump. The fact that we do not observe the LC bump
indicates that the helium mixing is not the only reason we observe
rapidly fading SNe~II. Even if the helium mixing plays a role in having a
large decline rate, other mechanisms like the \Ni\ mixing should be
accompanied to reduce the LC bump.

Finally, radii of SN~II progenitors may be significantly expanded,
as we discussed in Section~\ref{sec:radii}.
The expansion of the progenitor radii can result in large luminosity,
which makes the LC bump less significant,
and reduction of the hydrogen-rich envelope density (e.g., \citealt{mcley2014}),
which makes the LC decline more rapid.
In addition, the radius expansion of SNe~II progenitors with large
\Menv\
may also result in rapidly fading SNe~II with SN~IIP features like ASASSN-13co
\citep{holoien2014} because of the small envelope density with a large
hydrogen-rich envelope mass.
The observed rise times of SNe~II are suggested to indicate smaller progenitor radii for
more rapidly fading SNe~II, but the difference may not be significant
\citep{gall2015,gonzalez2015}.
The shock breakout observations as is recently reported by \citet{tominaga2015}
will further help to constrain the progenitor radii.

To summarize,
the LC bumps we find in our LC models for the progenitors with low-mass
hydrogen-rich envelopes which is not observed in SN~IIL LCs
indicates that rapidly fading SNe~II are not only characterized by
a low-mass hydrogen-rich envelope, which is responsible for the rapid LC decline,
but also by explosion properties, large degrees of \Ni\ mixing,
and/or large progenitor radii.
The later two properties, the large \Ni\ mixing and large progenitor
radii may also be a general characteristic of SNe~II including SNe~IIP,
but their consequences may simply be easier to recognize in rapidly fading SNe~II.

\section{Conclusions}\label{sec:conclusion}
We have presented our results of LC modeling of SNe~II from progenitors with
hydrogen-rich envelope masses near the solar mass.
We have confirmed that this kind of SN progenitors result in
rapidly fading SNe~II (SNe~IIL).
However, we find that theoretical LCs from such SN progenitors show a temporal
luminosity increase (LC bump)
shortly before all the hydrogen in the envelope recombines
when they have the canonical explosion energy of 1~B and \Ni\ mass of 0.05 \Msun.
The temporal luminosity increase is due to the heating from the \Ni\
decay, which keeps hydrogen ionized and make the photosphere stay at the
innermost layers of the hydrogen-rich envelope for a while.
Because this kind of temporal luminosity increase has not been observed,
rapidly fading SNe~II are likely to have
more properties in common than a low-mass hydrogen-rich envelope,
making the LC bump less significant.

We show four possible ways to make the LC bump less significant, i.e.,
\begin{enumerate}\raggedright
  \item large explosion energies ($\Eexp\gtrsim2$ B)
  \item small \Ni\ masses ($\Mni\lesssim 0.01~\Msun$)
  \item large degrees of \Ni\ mixing
  \item extended progenitor radii.
\end{enumerate}
We also find that the helium mixing into the hydrogen-rich layer does not
help reducing the LC bump, although it can make LC decline rates large.

Because rapidly fading SNe~II are observationally found to have larger
\Ni\ masses than slowly fading SNe~II \citep{anderson2014},
the lack of the LC bump in observed SN ~IIL LCs indicates that
the other three properties are likely to be shared
among SNe~IIL, in addition to the low-mass hydrogen-rich envelope.
If SN~IIL progenitors tend to have larger explosion energies than
those of SNe~IIP,
the empirical relation between the progenitor mass and the explosion
energy implies that SN~IIL progenitors are more massive than SN~IIP
progenitors.
Then, most of SN~IIL progenitors may not originate from small ZAMS mass
progenitors which can lose their hydrogen-rich
envelope by binary interactions.
Alternatively,
SNe~IIL may have larger degrees of \Ni\ mixing or more extended progenitor
radii than expected.

Another possible way to interpret the lack of the LC bump in SNe~IIL is
that not only SNe~IIL but
SNe~II including SNe~IIP may generally have large degrees of \Ni\ mixing and/or 
extended progenitor radii. The effect of large \Ni\ mixing and an extended
progenitor radius may be easier to recognize in
SN progenitors with low-mass hydrogen-rich envelopes through the LC bump
as we find in this study.
Constraining the degree of the \Ni\ mixing and the expansion of the progenitor
radius in SNe~IIL as well as SNe~IIP will be helpful to further constrain the intrinsic
properties of SNe~IIL.

\section*{Acknowledgments}
TJM is supported by Japan Society for the Promotion of
 Science Postdoctoral Fellowships for Research Abroad
 (26\textperiodcentered 51).
Pruzhinskaya~M.V. is supported by the Russian Foundation of Fundamental Research, grant RFFI 14-02-31546 and by Mechnikov Scholarship of the Embassy of France.
This work of S.~I. Blinnikov (development of \texttt{STELLA} code) was supported
by Russian Science Foundation grant 14-12-00203.
Numerical computations were partially carried out on Cray XC30 and PC cluster at Center for Computational Astrophysics, National Astronomical Observatory of Japan.
The Oskar Klein Centre is funded by the Swedish Research Council.
This research is also supported by World Premier International Research Center Initiative (WPI Initiative), MEXT, Japan.

\label{lastpage}

\end{document}